\documentclass{aa}
\usepackage{textcomp}
\usepackage{graphicx} %
\usepackage{amsmath} %
\usepackage{longtable} %
\usepackage{amssymb} %
\usepackage{bm} %
\usepackage{upgreek} %
\usepackage{IEEEtrantools} %
\usepackage{multirow}
\usepackage{booktabs}
\usepackage[dvipsnames]{xcolor}
\usepackage[colorlinks=true,allcolors=blue,urlcolor=blue,breaklinks=true]{hyperref}
\usepackage{txfonts}
\usepackage[normalem]{ulem} 
\usepackage{natbib}\usepackage{hyperref}
\hypersetup{
    colorlinks=true,
    linkcolor=blue,
    filecolor=blue,      
    urlcolor=blue,
    citecolor=blue
}

\bibpunct{(}{)}{;}{a}{}{,} 

\newcommand{\sect}[1]{\text{Sect.~\ref{#1}}}
\newcommand{\fig}[1]{\text{Fig.~\ref{#1}}}

\newcommand{\stagger}{\texttt{STAGGER}}
\newcommand{\cifist}{\texttt{CIFIST}}
\newcommand{\scate}{\texttt{Scate}}

\newcommand{\teff}{T_{\mathrm{eff}}}
\newcommand{\logg}{\log{g}}
\newcommand{\feh}{\mathrm{[Fe/H]}}
\newcommand{\dex}{\mathrm{dex}}
\newcommand{\gsun}{\ensuremath{g_{\odot}}}
\newcommand{\teffsun}{\ensuremath{T_{\text{eff},\odot}}}

\definecolor{cis}{rgb}{0.78, 0.08, 0.52}

\begin{document} 


\title{An extended and refined  grid of 3D \stagger{} model atmospheres}
\subtitle{Processed snapshots for stellar spectroscopy \thanks{\url{https://3dsim.oca.eu}}}

\author{Luisa~F.~Rodr\'iguez~D\'iaz\inst{\ref{aarhus}}
\and
Cis~Lagae\inst{\ref{su}}
\and
Anish~M.~Amarsi\inst{\ref{uu}}
\and
Lionel~Bigot\inst{\ref{oca}}
\and 
Yixiao~Zhou\inst{\ref{aarhus}}
\and
V\'ictor~Aguirre~B{\o}rsen-Koch\inst{\ref{dark}}
\and
Karin~Lind\inst{\ref{su}}
\and
Regner Trampedach\inst{\ref{ssi}}
\and 
Remo Collet\inst{\ref{aarhus}}}
\institute{\label{aarhus}Stellar Astrophysics Centre, Department of Physics and Astronomy, Aarhus University 
Ny Munkegade 120, DK-8000 Aarhus C, Denmark\\
\email{luisa-rodriguezdiaz@outlook.com}
\and
\label{su}Department of Astronomy, Stockholm University 
Albanova University Center, 106 91
Stockholm, Sweden
\and
\label{uu}Theoretical Astrophysics, 
Department of Physics and Astronomy,
Uppsala University, Box 516, SE-751 20 Uppsala, Sweden
\and
\label{oca}Université Côte d’Azur, Observatoire de la Côte d’Azur, 
CNRS, Lagrange UMR 7293, CS 34229, F-06304
Nice Cedex 4, France
\and
\label{dark}DARK, Niels Bohr Institute, University of Copenhagen
Jagtvej 128, 2200 
Copenhagen, Denmark
\and
\label{ssi}Space Science Institute
4750 Walnut Street, Suite 205
Boulder, CO 80301, USA
}

\abstract
    {Traditional one-dimensional (1D) hydrostatic model atmospheres introduce systematic modelling errors into spectroscopic analyses of FGK-type stars.}
  {We present an updated version of the \stagger{}-grid of 3D model atmospheres, and explore the accuracy of post-processing methods in preparation for spectral synthesis.}
   {New and old models were (re)computed following an updated workflow, including an updated opacity binning technique. Spectroscopic tests were performed in 3D LTE for a grid of 216 fictitious Fe I lines, spanning a wide range in oscillator strength, excitation potential and central wavelength, and eight model atmospheres that cover the stellar atmospheric parameter range ($\teff{}$, $\logg$, $\feh$) of FGK-type stars. Using this grid, the impact of vertical and horizontal resolution, and temporal sampling of model atmospheres on spectroscopic diagnostics was tested.}
  {We find that downsampling the horizontal mesh from its original size of $240\times240$ grid cells to $80\times80$ cells, i.e. sampling every third grid cell, introduces minimal errors on the equivalent width and normalized line flux across the line and stellar parameter space. Regarding temporal sampling, we find that sampling ten statistically independent snapshots is sufficient to accurately model the shape of spectral line profiles. For equivalent widths, a subsample consisting of only two snapshots is sufficient, introducing an abundance error of less than 0.015 dex.
  }
{We have computed 32 new model atmospheres and recomputed 116 old model atmospheres present in the original grid. The public release of the \stagger{}-grid contains 243 models, excluding models with $\feh$ = -4.00, and the processed snapshots can be used to improve the accuracy of spectroscopic analyses.}

\date{Received / Accepted: 13/05/2024}
\maketitle

\section{Introduction}
\label{introduction}

Three-dimensional (3D) compressible radiation-hydrodynamics simulations of stellar atmospheres, known as box-in-a-star type of model, represent the very essence of stellar atmosphere modeling and thereby spectroscopy for FGK-type stars. After the first studies on solar and stellar granulation and spectral line profiles \citep{Dravins1981, Nordlund1982, Nordlund1985, Dravins1990}, the field  evolved quickly with contributions in non-local thermodynamic equilibrium (non-LTE) spectroscopic analyses of the compositions of the Sun \citep{Nordlund1985,Kiselman1995,asplund00, Asplund2003a,Asplund2004,Caffau2008}. This led to, among other things, a significant revision of the solar chemical composition \citep{Asplund2009,Caffau2011b,Asplund2021}
that triggered a debate about the standard model of the Sun 
persisting to the present day \citep{2009ApJ...705L.123S,Christensen-Dalsgaard2021,Buldgen2023a,Buldgen2023b}.

In addition, 3D atmosphere models coupled with LTE and non-LTE radiative transfer have been applied to determine stellar parameters \citep{Amarsi2018,Bonifacio2018,BertranDeLis2022} and stellar abundances in metal-poor stars \citep[e.g.][]{Shchukina2005,Lind2013,Amarsi2016b,Nordlander2017,Bergemann2019,Lagae2023}.
Grids of 3D LTE \citep[e.g.][]{Chiavassa2018} and non-LTE spectra are now being calculated 
\citep[e.g][]{Amarsi2016,Mott2020,Wang2021} and
used to study Galactic chemical evolution
and Galactic archaeology for larger samples of stars \citep[e.g.][]{Sbordone2010,Amarsi2015,Amarsi2019,Giribaldi2021,Giribaldi2023}.
Continuum limb-darkening and centre-to-limb variations of spectral lines
have been studied in the Sun with 3D models
\citep[e.g.][]{Wedemeyer2009,Pereira2013,Lind2017}, and have been used 
to constrain atomic data \citep{Amarsi2018_osun,Amarsi2019_csun}. Furthermore, with the aid of these 3D models, interpolation becomes possible for spectra and limb darkening coefficients \citep[e.g.][]{Amarsi2018,Wang2021,BertranDeLis2022}, for equivalent widths and 
abundance corrections \citep[e.g.][]{Amarsi2016,Amarsi2019,Wang2021,Amarsi2022,Gallagher2016,Harutyunyan2018,Mott2020},
for photometric corrections \citep{Bonifacio2018,Chiavassa2018}, 
as well as for line shifts and convective blueshifts \citep{AllendePietro2013,Chiavassa2018}.

Recent studies have also extended the use of 3D model atmospheres to help with interpretation of
exoplanet transmission spectra \citep{flowers2019, Chiavassa2019,Canocchi2023}.
Moreover, 3D model atmospheres allow us to accurately model the granulation noise in stars \citep{Rodriguez2022}, provide an improved theoretical understanding of the properties (e.g. excitation, asymmetry) of pressure-mode oscillations \citep[e.g.][]{Nordlund2001,stein2001,georgobiani2003,samadi2001,samadi2003,philidet2020a,philidet2020b,belkacem2021}, as well as the so-called ‘‘surface effect''\citep{Houdek2017,Trampedach2017,Sonoi2019,Zhou2020,Zhou2021} and calibrate stellar structure models \citep{Trampedach2014a,Trampedach2014b}.

The main advantage of 3D model atmospheres over 1D models is that they do not need parameterised, approximate descriptions of the convective motions and their effects.
Convection arises naturally in 3D model atmospheres through heating from the bottom layers. Similarly, radiative cooling near the photosphere leads to the formation of a characteristic granulation pattern, consisting of hot upflows bounded by cool downdrafts.  In contrast, 1D models typically invoke mixing-length theory (MLT; \citealt{BoehmVitense1958}) or some variant of it \citep{Canuto1991} to attempt to describe energy transfer due to convection and reproduce temperature gradients in the star's atmospheres and influence the properties of spectral lines present in the atmosphere, to name a few examples. This theory cannot fully describe the broadening of spectral lines, nor line asymmetries or wavelength shifts, that we observe in stars and are attributed to stellar surface convection \citep{Dravins1981}. As such, spectroscopic analyses based on 1D models invoke various other ad hoc parameters, not least the so-called microturbulence and macroturbulence, which are assumed constants for a given atmosphere, but turn out to be line-dependent \citep[see more details in e.g.][]{Steffen2013, Jofre2014, Slumstrup2019}. In contrast, spectroscopy based on 3D modelling do not need to invoke these parameters \citep[e.g.][]{Asplund2000}.

An important result from 3D modelling 
is that the upper atmospheres in 3D models of low-metallicity stars are far cooler than predicted by 1D hydrostatic models in radiative-equilibrium \citep[e.g.][]{asplund99}. 
Inefficient radiative cooling in this regime means that the balance between radiative heating and adiabatic cooling is dominated by the latter, which cannot be predicted
by 1D models. This can translate to large effects for molecular lines in metal-poor stars, with abundance differences possibly reaching $1\,\dex$ \citep[e.g.][]{Collet2006,Collet2007,Gallagher2016}. 

The major downside of 3D models compared to 1D models is the increased computational complexity and cost. A symptom of this complexity are the various numerical parameters controlling the time step, numerical viscosity, grid size and resolution, and boundary conditions \citep{Asplund2000b,Grimm-Strele2015a,Collet2018}. The main approximation invoked to deal with the cost is 
the ''opacity binning'' or multi-group formalism \citep{Nordlund1982, Skartlien2000}, where the radiative heating and cooling rates are computed for just a handful of averaged opacities. 1D codes, on the other hand, can afford to perform a proper monochromatic opacity sampling, and solve the radiative transfer equation for hundreds of thousands or even millions of individual frequencies. Such a 1D calculation is also needed in order to assign individual wavelengths to bins in the opacity-binning method.

Driven by the groundbreaking results above yet hindered by the computational challenges, the construction of extensive grids of 3D model atmospheres is both a necessity and a priority to further advance the field. The \cifist{}-grid \citep{Ludwig2009,Tremblay2013} and the \stagger{}-grid \citep{Magic2013, Stein2024} stand in reflection of this.
Both have enabled 3D (non-)LTE spectroscopic analyses
of much larger samples of stars as referenced above.

In parallel to these fundamental developments in understanding and 
modeling of stellar atmospheres, the last 20 years
have seen a boom in large stellar surveys.
These include photometric missions like \texttt{CoRoT} 
\citep{Baglin2006}, \texttt{Kepler}
\citep{Borucki2010}, TESS \citep{Ricker2015}, and spectroscopic surveys 
such as RAVE \citep{Steinmetz2006}, LAMOST \citep{Cui2012}, GALAH 
\citep{DeSilva2015}, GAIA \citep{Gaia}, and  APOGEE \citep{Majewski2017}.
However, 3D model atmospheres have not seen much use in these surveys or in similar stellar surveys --- although a notable exception is GAIA Coordination Unit (CU6) \citep{GaiaDR1} and the recent application of the \stagger{}-grid to GALAH DR3 \citep{Wang2024}. For the first case, the GAIA/RVS pipeline implemented radial velocity corrections due to convective line-shifts \citep{Asplund2000,Chiavassa2018}, by using spectra computed with the \stagger{}-grid \citep{Bigot2008, Chiavassa2018}. 

With ever larger stellar surveys on the horizon including 4MOST \citep{deJong2019},
WEAVE \citep{Jin2023}, and PLATO \citep{Rauer2014}, there is a growing need for extensive grids of 3D model atmospheres.  
Here we present our efforts to refine and extend the \stagger{}-grid,
describing our workflow in detail (\sect{method}). 
We have recomputed 116 models of the original grid, and
have also computed 32 new models such that the final updated and refined version of the \stagger{}-grid contains 243 unique model atmospheres. These models are characterized by their effective temperature ($\teff$),
logarithmic surface gravity ($\logg$),
and logarithmic iron abundance relative to the sun ($\feh$\footnote{Defined as: $[\mathrm{A}/\mathrm{B}] = \log_{10}(N_\mathrm{A}/N_\mathrm{B})_\star - \log_{10}(N_\mathrm{A}/N_\mathrm{B})_\sun $, where $N_\mathrm{A}$ and $N_\mathrm{B}$ are the number densities of element A and B, respectively.}).
In addition, we discuss how to process these snapshots for stellar spectroscopy
applications (\sect{processing}),
namely by resampling horizontally and temporally, and by refining the vertical mesh.
For the first time, we make a grid of processed snapshots of 3D RHD models available to the community. \footnote{\href{https://3dsim.oca.eu}%
     	{https://3dsim.oca.eu}}
(\sect{conclusion}).

\section{Method}
\label{method}
\subsection{The \stagger{}-code}
\label{sec:staggercode}
In this section, we will briefly describe the \stagger{}-code. The \stagger{} compressible radiation-magnetohydrodynamic (R-MHD) code was originally developed by \cite{Nordlund1995} --- although the grid of models presented in this work does not include magnetic fields --- with more recent improvements by \cite{Magic2013} and \cite{Collet2018}, which can be consulted for further details. Additionally, a comprenhensive description of the code can be found in \citet{Stein2024}. \stagger{} solves the time-dependent hydrodynamic equations for the conservation of mass, momentum, and energy, together with the radiative transfer equation assuming LTE. The latter is solved along a total of nine rays, one vertical plus the combination of two inclined polar angles ($\mu = \cos{\theta}$) and four azimuthal angles ($\phi$-angles), in other words, eighteen directions on the unit sphere \citep{Stein2003}. 
Computing the full radiative transfer solution for all wavelengths and all grid cells, at every hydrodynamic timestep, is computationally unfeasible. Hence, \stagger{} employs the opacity binning method \citep{Nordlund1982, Skartlien2000, Ludwig2013, Collet2018} to approximate the full monochromatic problem. In summary, this method associates an average opacity strength to a subsample or 'bin' of wavelengths, selected according to their formation depth. The line opacities are taken from \cite{Gustafsson2008} and the sample of continuum absorption and scattering coefficients are taken from \cite{Hayek2010}. We use the Mihalas-Hummer-D{\"a}ppen (MHD) equation of state (EOS) \citep{Mihalas1988}, as implemented by \citep{Trampedach2013}. This EOS explicitly includes all ionization and excitation stages of all included elements, and a realistic treatment of plasma effects on bound states, via the so-called occupation probabilities. All the atomic physics is explicitly calculated for the required abundances, and is identical to that used for the original \stagger{}-grid.

\begin{figure}
    \centering
    \includegraphics[width=\columnwidth]{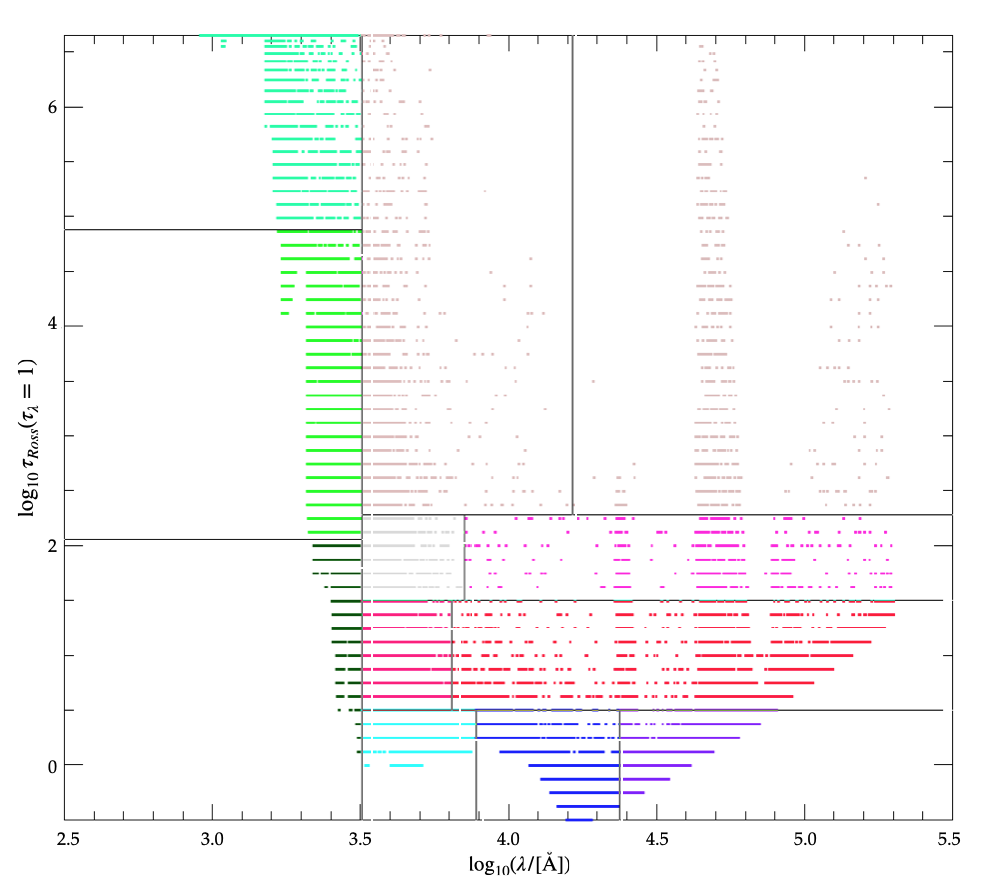}
    \caption{The improved distribution of the twelve opacity bins based on the results of \cite{Collet2018}. The Rosseland optical depth at the monochromatic photosphere, $\tau_\lambda=1$, in the average 3D ($\langle3\mathrm{D}\rangle$) reference atmosphere} is plotted against the wavelength, where each bin has been assigned a different colour. This representation is for a model with $\teff$=4500 K, $\log$=5.0 and $\feh$= +0.0.
    \label{fig:bins}
\end{figure}

In \citet[][Sec. 2.3.2]{Magic2013} it was discussed how the wavelengths are sorted in each bin, prioritising a fast and automatic selection of opacity bins for all models, despite the fact that they have different stellar parameters, especially metallicity. However, the distribution and limits of the bins directly affect the radiative heating and cooling, which in turn affects the temperature-density stratification of the simulation \citep{Collet2018}. This implies that the automated selection of bin sizes and their boundaries could have an important influence on the models. Therefore, we implemented a new distribution of the 12 bins, chosen because it best replicated the temperature stratification observed in the 48-bin configuration as shown in \cite{Collet2018}. An example of the new 12-bin distribution is shown in \fig{fig:bins}, where the bin limits are defined by the black lines and individually optimized for each model manually.

The differences between the original and the updated opacity binning can be seen in the temperature stratification in the upper atmosphere shown in Fig. 9 in \citet{Collet2018} and will be further discussed in Sec.~\ref{sec:running}.

\begin{figure}
    \centering
    \includegraphics[width=\columnwidth]{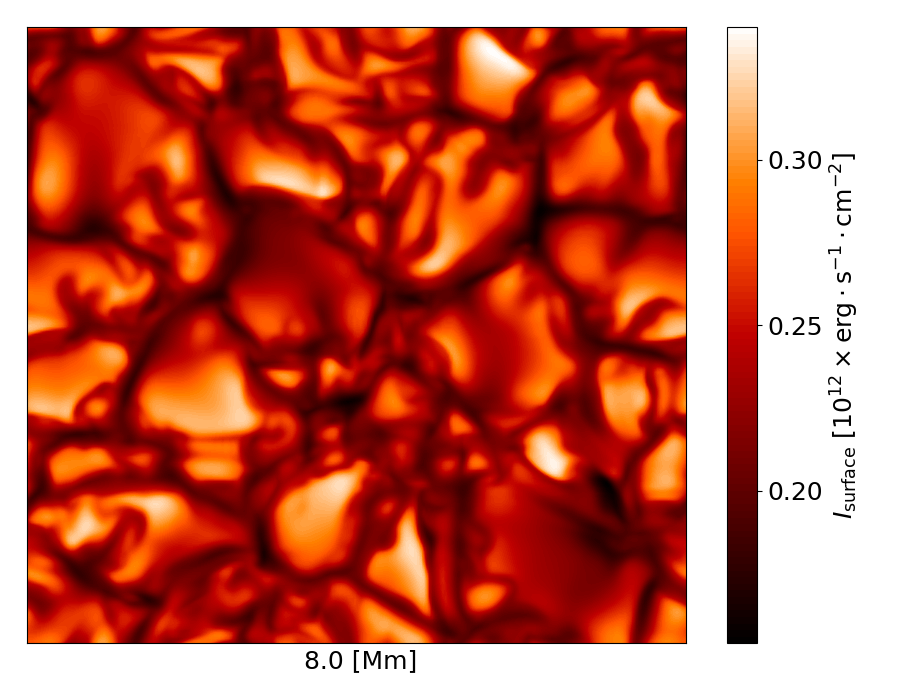}
    \caption{2D colour plot of a snapshot of the bolometric surface intensity of the solar model with dimensions $8\mathrm{Mm} \times 8\mathrm{Mm}$.}
    \label{fig:granulation}
\end{figure}

The 3D models generated with the \stagger{}-code are of box-in-a-star type, covering a small region of the stellar atmosphere. Specifically, the box includes the top layers of the convective zone, the super adiabatic region (SAR), the photosphere, and the upper atmosphere excluding the chromosphere. All models follow a Cartesian geometry with a constant gravity. The horizontal size is specified such that the box contains at least ten granules, see for example \fig{fig:granulation}. The horizontal boundary conditions are periodic, while the vertical boundaries are transmitting. Specifically, the inflowing gas at the bottom boundary has a constant value of specific entropy per unit mass $s$ which is specified through the values of density $\rho$, and internal energy per mass $\varepsilon$, of the inflows at the bottom ($\rho_{\rm bot}$ and $\varepsilon_{\rm bot}$). The value of $s_{\rm bot}(\rho_{\rm bot}$, $\varepsilon_{\rm bot})$ does not need to be known\footnote{The specific entropy is computed as an integration of $(\partial s/\partial\ln\rho)_\varepsilon=-p_{\rm gas}/(\rho T)$ along isochores in the EOS table, but is not needed for running simulations.}, but is maintained via the first law of thermodynamics:
\begin{equation}\label{eq:thermo_firststlaw}
    \mathrm{d}s=\frac{1}{T}\Big( \mathrm{d}\varepsilon - p_\mathrm{th}\frac{\mathrm{d}\rho}{\rho^2} \Big)~~,
\end{equation}
where $\rho$ is density and $\varepsilon$ the internal energy, and with the gas pressure $p_\mathrm{th}$ and temperature $T$ calculated from the equation of state. The internal energy per unit mass and density are user-specified and model-specific, and directly determine the emergent $T_\mathrm{eff}$ of the simulation resulting from the complex, non-linear interactions between hydrodynamics, radiative transfer and thermodynamics.

\subsection{Initializing and running a \stagger{} model}
\subsubsection{Initializing a new model}
\label{sec:newmodel}
The procedure to start a new 3D stellar atmosphere model is based on similar principles as described in \citet[see Sec. 2.3.1]{Magic2013}. The goal is to produce an initial snapshot that is as close to the final, new, relaxed simulation as possible, balanced with requirements of a robust method that is generally applicable for any late-type stellar atmospheric parameters. This is accomplished with a plane-parallel atmospheric model with adiabatic stratification in (quasi-) hydrostatic equilibrium. 

To do so, a relaxed snapshot from another existing 3D model with equal $\feh$ and close $\teff$ and $\logg$ is used, which we will refer to as the reference model. The key point is, that the reference model should have a similar specific entropy per unit mass at the bottom boundary, as this will determine the emergent $\teff$ of the model. Moreover, a suitable 3D reference model will allow for a better and more efficient scaling and relaxation of a new model \citep{Magic2013}. Due to the adiabatic nature of the stellar sub-surface convection, the mean entropy in these layers is approximately constant \cite{Steffen1993}. Subsequently, from the EOS and the given entropy value at the bottom boundary, an adiabatic density and pressure stratification can be computed, which is in turn used to determine a geometrical depth scale that fulfills hydrostatic equilibrium. This results in an initial guess for the profiles of the main thermodynamic variables and the geometrical depth scale of the new 3D model, based on the reference model. In addition, the opacity binning scheme of the reference model is copied for the new model, which will be refined in a later step (see \sect{sec:running}). The final initial snapshot is then a plane-parallel atmospheric model with adiabatic stratification in (quasi-) hydrostatic equilibrium.

The horizontal dimensions of the model \emph{l$\times$l} are calculated as follows:
\begin{equation}
    l = l_\odot\left(\frac{\gsun}{g}\right)\left(\frac{\teff}{\teffsun}\right)~~,
\end{equation}
where $l_\odot$ is the horizontal dimension of the Solar model. In other words, we rescale the horizontal dimensions by an approximate, surface pressure scale height, at constant metallicity. These dimensions are sufficient to capture at least ten granules at any given time. The vertical extent of a new 3D model is chosen to be on the order of ten times the surface pressure scale height and is set on a uniform mesh. 

\subsubsection{Relaxing a model}
\label{sec:running}
Allowing the simulation to settle into its dynamic, but quasi-static 'natural' state, is known as the relaxation of the simulation. In addition, trying to impose a model with a predefined (target) value of the effective temperature demands some adjustments and more iterations since $T_{\rm eff}$ is an output of the simulation. Hence, a major part of the relaxation process is spent on adjusting the bottom boundary conditions to achieve the desired effective temperature.
The procedure outlined in this work is build upon the original workflow as described in Sec 2.3 in \cite{Magic2013}.

{\it Step 1}: First, we run the initial snapshot (at `low' grid resolution $120^3$) for multiple turnover times with only one vertical ray to compute the radiative heating and cooling rates. We frequently save snapshots of the model, roughly every $H_p/c$ time, which is the ratio of surface pressure scale height and the sound speed. For the Sun, this value is approximately 20s. This quick first run essentially checks that the simulation allows convective motions to emerge and does not evolve away from quasi-hydrostatic equilibrium. 

{\it Step 2}: We increase the number of rays used in the radiative transfer solver to three $\mu$ = cos$\theta$ angles (two inclined and one vertical) and four $\phi$-angles. This new setup provides a more accurate representation of radiative heating and cooling in the atmosphere, without being too computationally expensive \cite{Stein2003}. After this change, we run the simulation for at least three turnover times. Since the radiative transfer is now more accurately treated, the model has a $T_{\rm eff}$ that is close to its true $T_{\rm eff}$, where true means the one corresponding to the injected heat flux from below (internal energy, density), the gravity and the metallicity.

{\it Step 3}: At this point, we check if the $T_{\rm eff}$ of the model is reasonably close to the target $T_{\rm eff}$. If yes, then no change of the incoming heat flux is needed. Otherwise, we slightly adjust the entropy at the bottom by changing the value of either the in-flowing internal energy per unit mass or density at the lower boundary. If the initial model was correctly defined, the change of those quantities must be small and thereby we assume that running the model several turnover times is enough to obtain the thermal relaxation. The Kelvin-Helmoltz time that describes this timescale is longer than the turnover time; typically for the Sun it is roughly a few hours at the depth that corresponds to the bottom of our simulations \citep[e.g.][]{Kupka2017}. Therefore, running a long sequence with many turnover times ($\geq 10-20$) ensures to have the thermal equilibrium. In the whole workflow as described in this subsection we reach such duration. The thermal relaxation is nonetheless checked by looking if there is no drift of the effective temperature (Fig 4), see also \sect{sec:relaxation}. 

\begin{figure}
    \centering
    \includegraphics[width=\columnwidth]{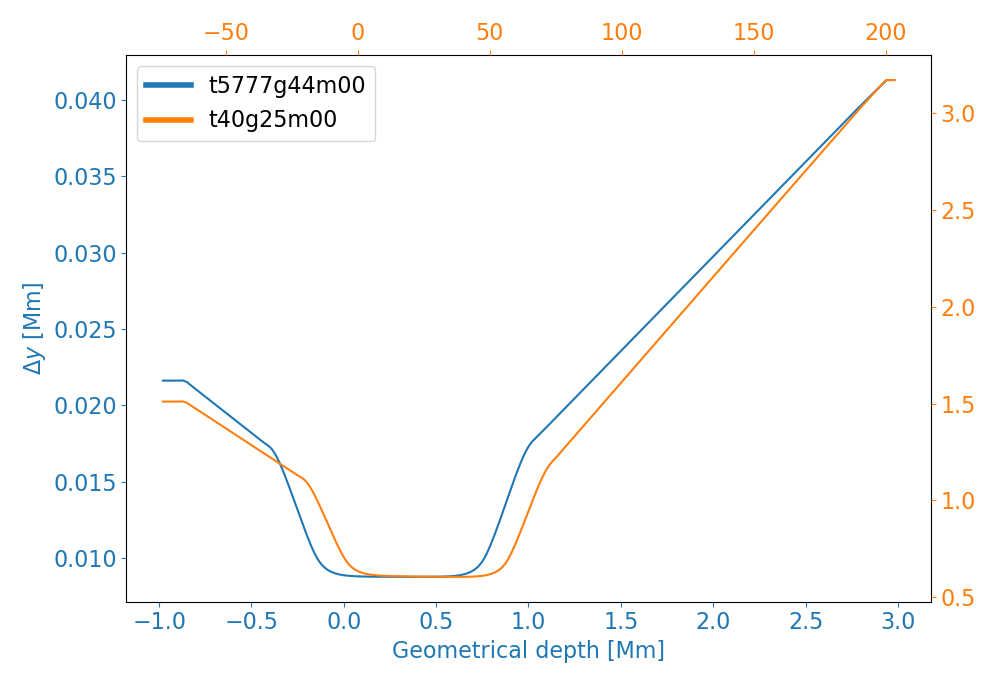}
    \caption{Example of the non-uniform vertical spacing of grid cells for the solar model $\teff$=5777 K, $\log$=4.4 and $\feh$= +0.0, and a cool Giant $\teff$=4000 K, $\log$=2.5 and $\feh$= +0.0.}
    \label{fig:GridRefinement}
\end{figure}

{\it Step 4}: We proceed to refine the geometrical depth scale. Initially, the grid points are uniformly distributed along the vertical direction, but to resolve the steep temperature gradients in the SAR, at the moderate mesh resolution used for the \stagger{}-grid, more grid points are needed in this region. The idea behind this method is to distribute the mesh points as evenly as possible on the optical-depth scale \citep{Magic2013}. We suggest the reader consult Sec. 2.3.1 of \citet{Magic2013} for details about this step. In contrast to their approach, we no longer automate this step. Instead, the process to refine the geometrical depth scale is done individually and manually for every model, to properly follow the atmospheric stratification of the model, which varies greatly with stellar parameters. An example of a main sequence model and a model in the red giant branch is shown in \fig{fig:GridRefinement}.

{\it Step 5}: To further improve the model, the opacity binning scheme is updated. The overall bin distribution is kept the same, but individual bin boundaries are adjusted as the simulation has reached a new thermal state as compared to the initial snapshot. As mentioned in Sec. 2.1, the opacity binning method is implemented to distribute wavelengths into a small number of groups of bins, each with its own opacity and source function. 

To do so, we compute the mean density and temperature stratifications of the model atmosphere, carry out a full monochromatic radiative transfer calculation as a reference case and perform an optimization of the bin limits based on the obtained monochromatic heating rates. The optimization is done for each bin limit individually, where we choose the solution (i.e. bin boundary location) that results in the smallest relative difference between the total heating rates as a result of opacity binning and the full monochromatic solution, as mentioned in \citep[][see their Eq. 13]{Magic2013}. To obtain the optimal configuration, we iterate over all bin limits until the position of the bin limits stop changing.\\

{\it Step  6}: Convective fluctuations excite acoustic modes naturally in the simulation as it does in real stars.  However, extra acoustic modes are generated during the relaxation process. They arise as a response to changes made to the model (mesh, inflowing internal energy) which in turn adjusts to converge towards its new hydrostatic equilibrium by generating extra pressure perturbations. Since our simulations are very shallow, these acoustic modes have low inertia and therefore can have sufficiently large amplitudes that can crash the simulations. They need to be damped during the relaxation process. We achieve this by damping the vertical mass flux in the following way:
\begin{equation}
    \partial_t  p_y (x,y,z) \rightarrow \partial_t p_y (x,y,z,t) - \rho (x,y,z,t) \frac{\langle p_y\rangle}{\langle\rho\rangle t_{\rm damp}}~~,
\end{equation}
where $\langle...\rangle$ are horizontal averages, $p_y$ the vertical component of momentum density and $t_{\rm damp}$ is the damping time which corresponds to the period of the fundamental mode. The value depends on the star ($T_{\rm eff}$, $\log g$, [Fe/H]) and the vertical extend of the simulation box, and is computed from the Fourier spectrum of the time-series of the vertical mass flux during the relaxation stage.

Then, we proceed to decrease the damping by twice and four times the original value – each time running for only one convective turnover time – so that the box modes of the simulation can emerge at their `natural' amplitude (i.e. the amplitude of the modes of the relax model). Next, we switch the damping off completely and let the simulation run for two convective turnover times. 
\\
In the last step, we increase the number of grid points from $120$ to $240$ in all directions. We run again the model for several convective turnover times, update the opacity table and once again damp the box modes. Finally, we initiate the final run of the fully relaxed model, without mode damping, which extends for at least three convective turnover times.

\subsection{Criteria for relaxation}\label{sec:relaxation}

\begin{figure}
    \centering
    \includegraphics[width=\columnwidth]{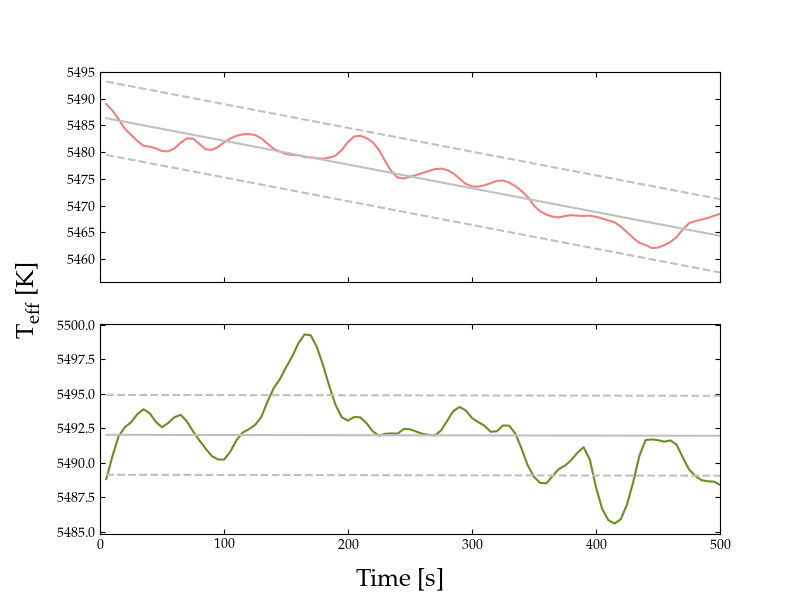}
    \caption{Time evolution of $\teff$ for the same model with $\teff$ = 5500 k, $\logg$ = 5.0, $\feh$ = -4.0. The top panel shows the original non-relaxed model from the grid, with $\teff$ decreasing quite quickly, while the new relaxed model is shown in the bottom panel. The solid grey lines indicate the mean $\teff$ at a given time step, while the dashed lines are spaced out by 1-$\sigma$ (standard deviation).}
    \label{fig:ComparisonRelaxationTeff}
\end{figure}

In order to check whether a simulation is relaxed, we defined three methods based on earlier work by \citet{Magic2013} and \citet[Sec. 2.4,][]{Trampedach2013}. First, we check the temporal evolution of the mean bottom boundary internal energy per unit mass and density, effective temperature and box-mode oscillations. While the bottom boundary internal energy per unit mass and density of the upflows are fixed, the mean value, including both down- and upflows of these quantities at the bottom boundary, can vary slightly due to the convective motions present in the model. If the model is relaxed, the mean bottom boundary internal energy per unit mass and density should fluctuate around a constant value over time. Similarly, the emergent effective temperature should not display a gradient with respect to time. These quick but important checks can immediately show if a model is relaxed or not. An example of this is shown in \fig{fig:ComparisonRelaxationTeff}, where the top panel shows a non-relaxed model, while the bottom one shows a new relaxed version of the same model.

Secondly, we inspect the temporal evolution of several horizontally averaged quantities of a simulation. These include temperature, density, internal energy per unit volume, vertical and root-mean-square (rms) velocity. These horizontally-averaged quantities are computed on layers of constant geometrical and optical depth for all individual snapshots ($\approx 300$) of the final run, covering three convective turnover times, following the methodology described in \sect{sec:temporalsampling}. Finally, these quantities are averaged over all snapshots creating a temporal mean of the full-time series. If the model is relaxed, the horizontally averaged quantities of all individual snapshots should be consistent with that of the full-time series and there should be no clear trend with respect to time. The latter is checked by comparing the horizontally averaged quantities of the first 20 snapshots to the last 20 snapshots of the simulation.

The third and final step is to inspect the horizontally averaged entropy, to verify that it has a constant value with depth in the convective zone, and verify that hydrostatic equilibrium is fulfilled.


\section{An extended and refined \stagger-grid} \label{sec:newgrid}
\subsection{Description of the grid}
The original \stagger{}-grid computed by \cite{Magic2013} consisted of 217 models covering a broad range of stellar effective temperatures and surface gravities for seven metallicities. Since its production, the grid has been widely used for several applications, which is a testament of the usefulness, importance, and need of this grid for the astronomical community. Nonetheless, based on the criteria for relaxation discussed in the previous section, some deficiencies were found, namely multiple model atmospheres showed signs of non-relaxation and a few were missing from the grid. Motivated by this and the need of a grid with more coverage, we have worked on a new version of the \stagger{}-grid, with fully converged models.
 
For the majority of the flawed models, relaxation was achieved by running the model for several more convective turnover timescales. Some models required small adjustments to the bottom boundary internal energy, to obtain an effective temperature closer to the target value. Wherever possible, missing models were computed from scratch. Furthermore, we aimed to refine the grid by adding 32 models in the parameter space of particular relevance for upcoming space mission such as PLATO:
\begin{itemize}
    \item $\teff$ = 3500 K, $\logg$ = 4.50, $\feh$ = 0.0
    \item $\teff$ = 4000 K, $\logg$ = 4.50, $\feh$ = 0.0
    \item $\teff$ = 4750 K, $\logg$ = 3.25, $\feh$ = 0.0
    \item $\teff$ = 4750 K, $\logg$ = 4.25, all seven metallicities 
    \item $\teff$ = 4750 K, $\logg$ = 4.75, all seven metallicities 
    \item $\teff$ = 5250 K, $\logg$ = 4.25, all seven metallicities 
    \item $\teff$ = 5250 K, $\logg$ = 4.75, all seven metallicities 
    \item $\teff$ = 6000 K, $\logg$ = 5.00, $\feh$ = 0.0
\end{itemize}

\begin{figure}
    \centering
    \includegraphics[width = \columnwidth]{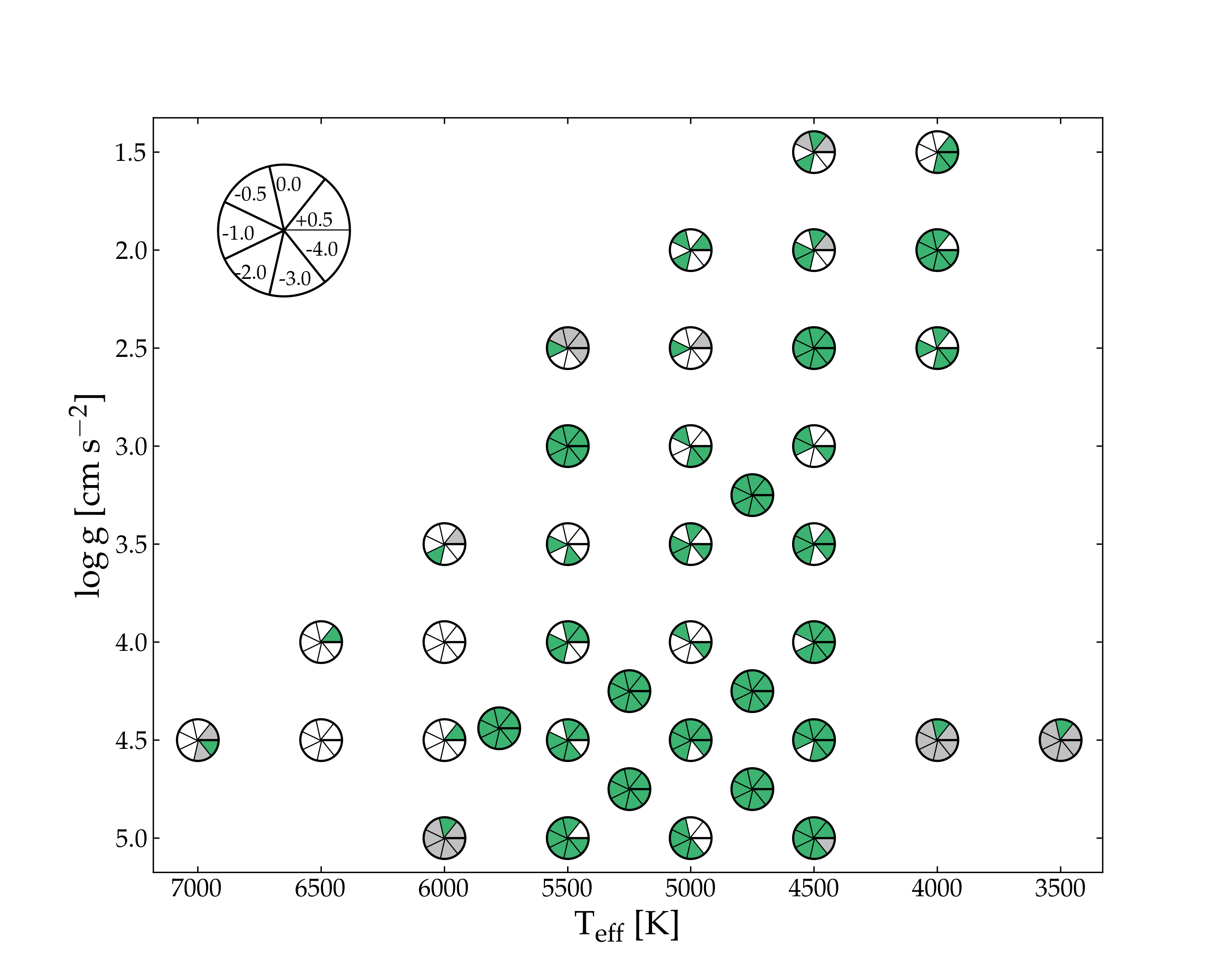}
    \caption{Status of the new and updated \stagger-grid, where green indicates the new/updated models, grey missing/not planned models, and white represents the unchanged models from the original grid.}
    \label{fig:newgrid}
\end{figure}

The result of this work is a new extended and refined grid of 3D model atmospheres containing 243 models, excluding models with $\feh$ = -4.00, as will be explained in \sect{sec:m400}. 
\fig{fig:newgrid} shows an overview of the new grid, where the new and updated models are in green (148),  missing/not planned models in grey, and white represents the unchanged models from the original grid.

The published grid will contain ten snapshots for each model atmosphere, selected using the procedure defined in \sect{sec:temporalsampling}, both in their original format and a processed format that is ready to be used in spectrum synthesis codes. The latter will be discussed in more detail in \sect{processing}. Every snapshot contains information on gas temperature, density, internal energy and momentum at each grid point, corresponding to 316 MB or 35 MB memory for both formats, respectively. The mesh information (size and spacing) are stored in two separate files that are provided together with the snapshots. In addition, the equation of state is stored in a separate binary file, which can be used to compute additional variables such as pressure and optical depth at $500$ nm. All snapshots, mesh files and complementary analysis scripts are accessible from the new \stagger{} website\footnote{\stagger{}-website: \href{https://3dsim.oca.eu}%
     	{https://3dsim.oca.eu}}.

\subsection{Flagging of peculiar models}
\label{sec:chromosphere}

In this study, we identified peculiar temperature structures in the outermost layers of both old and new model atmospheres. Specifically, we observed an outward increase of the geometrically- and temporally-averaged temperatures of several hundred kelvin in the optical thin layers ($\log\tau_\mathrm{500}\lessapprox-4$) of 31 models. These models are not limited to a certain region of the HR-diagram and range from solar to ultra metal poor ($\mathrm{[Fe/H]}=-4.0$). Some heating of the upper atmospheric layers is expected through the energy deposition of acoustic waves generated at the surface by convective motions. Combined with the magnetic fields in the Sun, this gives rise to magneto-acoustic heating of the chromosphere, which is well studied for the Sun \citep{Carlsson1992,Carlsson1995,Fawzy2002,Abbasvand2020,Murawski2020,Wojcik2020}, but has gotten limited attention for other stars. For example, \citet{Wedemeyer2017} computed a vertically extended 3D model atmosphere of a red giant star, omitting magnetic fields, using \texttt{CO5BOLD}. Their model developed a highly dynamical chromosphere characterized by acoustic waves evolving into strong shock fronts, heating the gas up to 5000 K. However, these hot regions were not represented in the geometrically averaged temperature stratification, which did not show a significant temperature increase in the chromospheric layers. We observe similar behaviour in the majority of our models that extend into higher atmospheric layers ($-4>\log\tau_\mathrm{500}\gtrapprox-10$), i.e. acoustic waves heating up the gas locally up to $5000 - 6000~\mathrm{K}$ without a corresponding increase in the mean temperature stratification.

Regarding the 31 models that do show a significant increase in the mean temperature profile, further investigation is required to understand if these layers are a physical or numerical artefact, for example due to a badly-constructed opacity binning tables. Regardless, since these `chromospheric' structures occur at small optical depths, we expect no significant impact for most spectroscopic applications such as photospheric line formation. Specifically, for several individual models it was already seen that these layers do not impact spectral line formation and, hence, could be ignored in subsequent abundance analysis \citep{Lagae2023}. These extremely optically-thin layers typically only impact the cores of the strongest lines. Weak metal lines (key diagnostics for chemical compositions, as well as for stellar parameters via excitation and ionisation equilibrium) and the wings of strong lines (important diagnostics of effective temperature and surface gravity) are unaffected. As such, we proceed to release these models in tandem with the rest of the grid with an added flag (\texttt{pec$\_$outer}) to inform users.

\section{Processed snapshots}
\label{processing}

\subsection{Overview}
A major hurdle in performing spectrum synthesis in 3D and non-LTE is the computational cost, both computing time and memory requirement. The cost of such calculations scales with the size of the model atmosphere $N_\mathrm{x}N_\mathrm{y}N_\mathrm{z}$ times the number of rays $N_\mathrm{ray}$, number of frequencies $N_\nu$ and number of iterations $N_\mathrm{iter}$. A simple estimate shows that the computational time of 3D non-LTE spectrum synthesis is at least a factor $10^5$ larger than corresponding 1D non-LTE calculations \citep{Nordlander2017,Asplund2021, Lind2024}. In addition, it is necessary to compute a spectrum for several snapshots of the hydrodynamic simulation to capture the spatial variation of the convective stellar surface. Therefore, it is desirable to post-process the model atmospheres to reduce the computational cost without sacrificing in accuracy. In the next subsection, we will go into the details of three methods to post-process model atmospheres and test their corresponding impact on spectral diagnostics using the 3D LTE radiative transfer code \scate{} \citep{Hayek2011}.

\subsection{Trimming}\label{sec:trimming}
The deep layers of the model atmosphere where the the radiation field satisfies $I_{\nu}\approx B_{\nu}$ do not contribute to the spectral line formation but do add to the computational cost. Therefore, we can safely cut away these layers prior to preforming the radiative transfer calculations. Previous tests to ensure this has minimal effect on the line formation have been preformed by \cite{Asplund2000}, \cite{Asplund2000b}, \cite{Asplund2003a}, \cite{Collet2006} and \cite{Amarsi2016}. In this work we aimed to reduce the vertical extent of the models to a size similar to \texttt{MARCS} atmospheres: $-5<\log\tau<2$ \citep{Gustafsson2008}. However, since every column in the 3D model atmosphere has a different optical depth scale $\tau(z)$, we remove all layers with geometrical height z that satisfy:
\begin{equation}
    -5>\max\big(\log\tau(z)\big)~~
\end{equation}
and
\begin{equation}
    ~2<\min\big(\log\tau(z)\big)~~.
\end{equation}
The resulting atmosphere is then interpolated onto a new grid that is more refined near the optical surface of the star based on the vertical gradient of $T^4$. This new depth scale will allow for better accuracy in subsequent radiative transfer calculations and spectral synthesis. As an example, for a random snapshot of the solar simulation, with size $240\times240\times240$, this procedure removed $118$ layers from the bottom and zero layers from the top of the model atmosphere. The remaining $122$ layers are then interpolated onto the new depth scale, with a user-specified number of grid points. In the next section, we investigate the effect of interpolating the trimmed atmosphere on the new depth scale, but with varying number of grid points: 240, 120 and 80.

\subsection{Horizontal and vertical grid sampling}\label{sec:Hrefinement}
The high resolution \stagger{} models are set on a cubic mesh of 240 cells. Such high resolution is needed in order to resolve features in the simulations when running them, and is therefore crucial for the reliability of any post-processing calculations \citep{Asplund2000b}. With the simulations already established, however, computationally expensive post-processing, such as spectral synthesis, can greatly benefit from down-sampling the simulations, with little loss of accuracy. For example, 3D non-LTE spectrum synthesis with full-resolution stagger models is unfeasible for large atoms such as Fe due to the excessive computational cost and memory requirements. One possible solution to reduce the computational cost, is to downsample the horizontal and vertical mesh while retaining the physical extent of the model atmosphere. This method has been extensively used in previous spectrum synthesis calculations, both 3D LTE \citep{Asplund2000,Asplund2001,Collet2006,Frebel2008,Kucinskas2013} and 3D non-LTE \citep{Asplund2003a,Amarsi2015,Steffen2015,Amarsi2016,Bergemann2019,Bergemann2021, Nordlander2017,Wang2021, Lagae2023}. Specifically, \cite{Nordlander2017} downsampled a single relaxed \stagger{} model to a range of resolutions which were subsequently used in 3D non-LTE spectrum synthesis. They found that downsampling the horizontal mesh of the model atmosphere from $240^2$ to $60^2$ grid points introduced an error on the equivalent width of less than $0.03~\dex$. \cite{Steffen2015} found that downsampling the horizontal direction of their 3D solar model by a factor three resulted in a flux difference less than $0.1\%$ and equivalent width error of $\delta W_\lambda/W_\lambda\approx12\times10^{-4}$ for the \ion{O}{I} IR triplet. Similarly, \cite{Bergemann2021} studied the effect of horizontal resolution of the solar \ion{O}{I} lines in 3D LTE for both a \stagger{} and \texttt{bifrost} model atmosphere. They downsampled the original horizontal mesh of size $240^2$ to $5^2$, $10^2$, $20^2$, $30^2$ and $120^2$, yielding flux errors less than $2\%$ and abundance errors of $0.01~\dex$ in the case of $30^2$.

In this work, we build on these previous studies by investigating the effect of horizontal and vertical resolution over a large parameter space. We synthesized a grid of 216 fictitious Fe I lines in 3D LTE covering: $\log{gf}=0,-1,-2,-3,-4,-5~\mathrm{dex}
$, excitation potential $E_l=0-5~\mathrm{eV}~~(\Delta=1\mathrm{eV})$ and wavelength $\lambda=300-1300~\mathrm{nm}~~(\Delta=200~\mathrm{nm})$. This grid was computed using \scate{} for nine different \stagger{} models (Table \ref{t:stagger parameters}) that were trimmed according to \sect{sec:trimming}, and subsequently interpolated to six spatial resolutions (Table \ref{t:stagger resolutions}). The Fe abundance was set to solar \citep{Asplund2021} or solar scaled $[\mathrm{Fe/H}]=-2$, depending on the model atmosphere. In these tests we treat the trimmed and interpolated high-resolution \stagger{} model, with size $240^2\times240$, as the reference model. To quantify the impact of reducing the horizontal and vertical resolution on the spectral synthesis, we compared the normalized flux and equivalent width of the lower resolution models to the high resolution reference model. We limit this analysis to lines that lie on the linear part of the curve of growth:
\begin{equation}
    -6.5<\log\big(\frac{W_\lambda}{\lambda}\big)<-5~~,
\end{equation}
which we will refer to as weak lines; and those lying on the flat part of the curve of growth: 
\begin{equation}
    -5<\log\big(\frac{W_\lambda}{\lambda}\big)<-4.5~~,
\end{equation}
which we will refer to as saturated lines in the rest of the paper.

\begin{table}[t]
\begin{center}
    \caption{Summary of the \stagger{} models used in the spatial sampling tests.}
    \def\arraystretch{1.5}
    \begin{tabular}{ cccc}
     \hline
     \hline
      Name & $T_\mathrm{\!eff}$ [K] & $\log g$ [dex] & [Fe/H]\\ \hline
      t5777g44m00 & $5775 \pm 12$ & $4.4$ & $0$ \\
      t5777g44m20 & $5775 \pm 9$ & $4.4$ & $-2$ \\
      t45g20m00 & $4490 \pm 10$ & $2.0$ & $0$ \\   
      t45g20m20 & $4480 \pm 10$ & $2.0$ & $-2$ \\   
      t50g20m20 & $4990 \pm 10$ & $2.0$ & $-2$ \\
      t45g50m00 & $4507 \pm 6$ & $5.0$ & $0$ \\
      t45g50m20 & $4501 \pm 1$ & $5.0$ & $-2$ \\
      t65g40m00 & $6430 \pm 15$ & $4.0$ & $0$ \\
      t65g40m20 & $6440 \pm 15$ & $4.0$ & $-2$ \\
      
     \hline
    \end{tabular}
    
\label{t:stagger parameters}
\end{center}
\end{table}

\begin{table}[t]
\begin{center}
    \caption{Summary of the grid sizes used in the spatial sampling tests.}
    \def\arraystretch{1.5}
    \begin{tabular}{ccc}
     \hline
     \hline 
      Name & $N_\mathrm{x}$, $N_\mathrm{y}$ & $N_\mathrm{z}$ \\ \hline
      $240^2\times240$ & $240$ & $240$  \\
      $240^2\times120$ & $240$ & $120$   \\
      $240^2\times80$ & $240$ & $80$    \\
      $120^2\times240$ & $120$ & $240$  \\
      $80^2\times240$ & $80$ & $240$    \\
      $80^2\times120$ & $80$ & $120$    \\
     \hline
    \end{tabular}
\label{t:stagger resolutions}
\end{center}
\end{table}

We quantify the impact of the lower resolution models on the equivalent width by taking the log of the ratio: 
\begin{equation}
    \log \frac{W_\lambda}{W_{\lambda,\mathrm{ref}}} \sim \log A - \log A_\mathrm{ref}~~,
\end{equation}
as this directly translates into an abundance difference for lines on the linear part of the curve of growth. For the saturated lines, we constructed curve of growths for a single line ($\lambda=5000~\AA$, $E_l=2~\mathrm{eV}$, $\log gf=-4$) for all model atmospheres in Table \ref{t:stagger parameters} and performed linear regression on the flat part (i.e. $-5<W/\lambda<-4.5$). Since the curve of growth varies depending on the stellar parameters, we adopt the worst case as a uniform relationship between equivalent width and abundance, for all stellar parameters and spectral lines:
\begin{equation}\label{eq:satlines}
    \log \frac{W_\lambda}{W_{\lambda,\mathrm{ref}}} \sim 0.2\cdot\Big(\log A - \log A_\mathrm{ref}\Big)~~.
\end{equation}

\begin{figure*}[ht]
    \centering
    \includegraphics[width=0.96\textwidth]{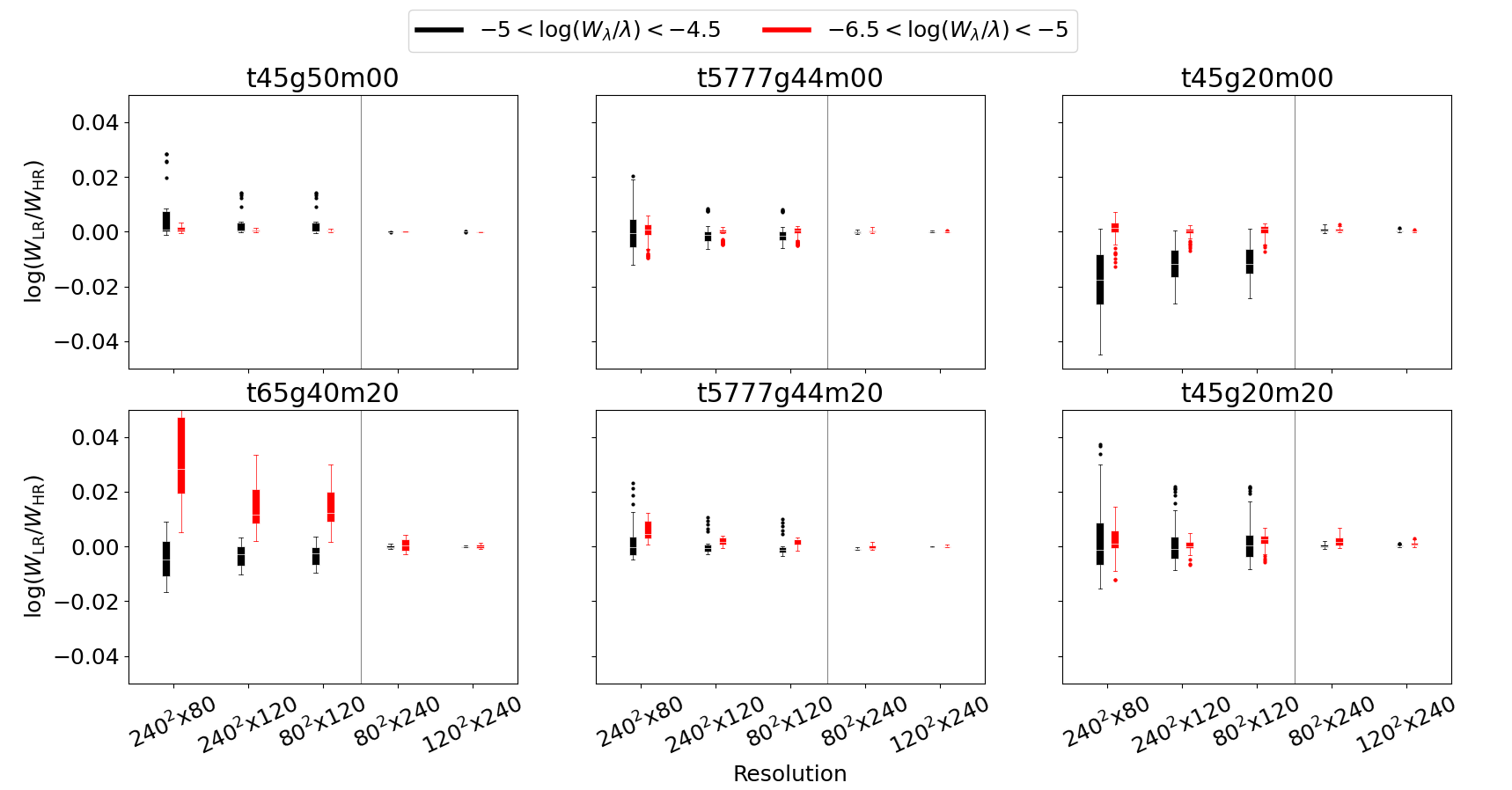}
    \caption{Overview of the relative difference in reduced equivalent width for five different spatial resolutions compared to the reference $240^2\times240$ resolution, computed from the 3D LTE flux spectrum. Each boxplot contains information from all lines that satisfy the reduced equivalent width requirement shown in the legend. The median of the boxplot is represented by a horizontal white line and outliers with a black or red dot. The vertical grey line divides the models with reduced vertical and horizontal resolution. }
    \label{fig:EW_boxplots}
\end{figure*}

The results are visualized in \fig{fig:EW_boxplots} using boxplots for both the saturated and weak lines separately, and six model atmospheres. The box size is set by the first $q1$ and third $q3$ quartile, containing 50$\%$ of all data points centred around the median. The two whiskers each extend to the smallest and largest data point that are not outliers, with outliers defined as points that are at a distance of $>1.5\times(q3-q1)$ from the box. 

From \fig{fig:EW_boxplots}, we deduce that in general, saturated lines show a larger dependency on resolution. Note that for the saturated lines the abundance error has to be multiplied by a factor five, see Eq. \ref{eq:satlines}. However, weak lines become more sensitive to resolution for models at lower metallicity, while the opposite is true for saturated lines. In addition, the error on equivalent width increases with decreasing surface gravity and increasing stellar effective temperature, seen by the large abundance errors for t65g40m20 and t45g20m00. Regarding only weak lines, downsampling the vertical mesh to 120 grid points results in an abundance error of less than $0.02~\dex$ for cool dwarfs and giants. For hotter models such as t65g40m20, the abundance error is at most $0.03~\dex$. On the other hand, downsampling the horizontal mesh seems to have minimal impact on the equivalent widths with abundance errors less than $0.007~\dex$. These results are in line what has been observed by \cite{Steffen2015}, \cite{Nordlander2017}, \cite{Amarsi2017} and \cite{Bergemann2021}.

\begin{figure*}[h]
    \centering
    \includegraphics[width=0.96\textwidth]{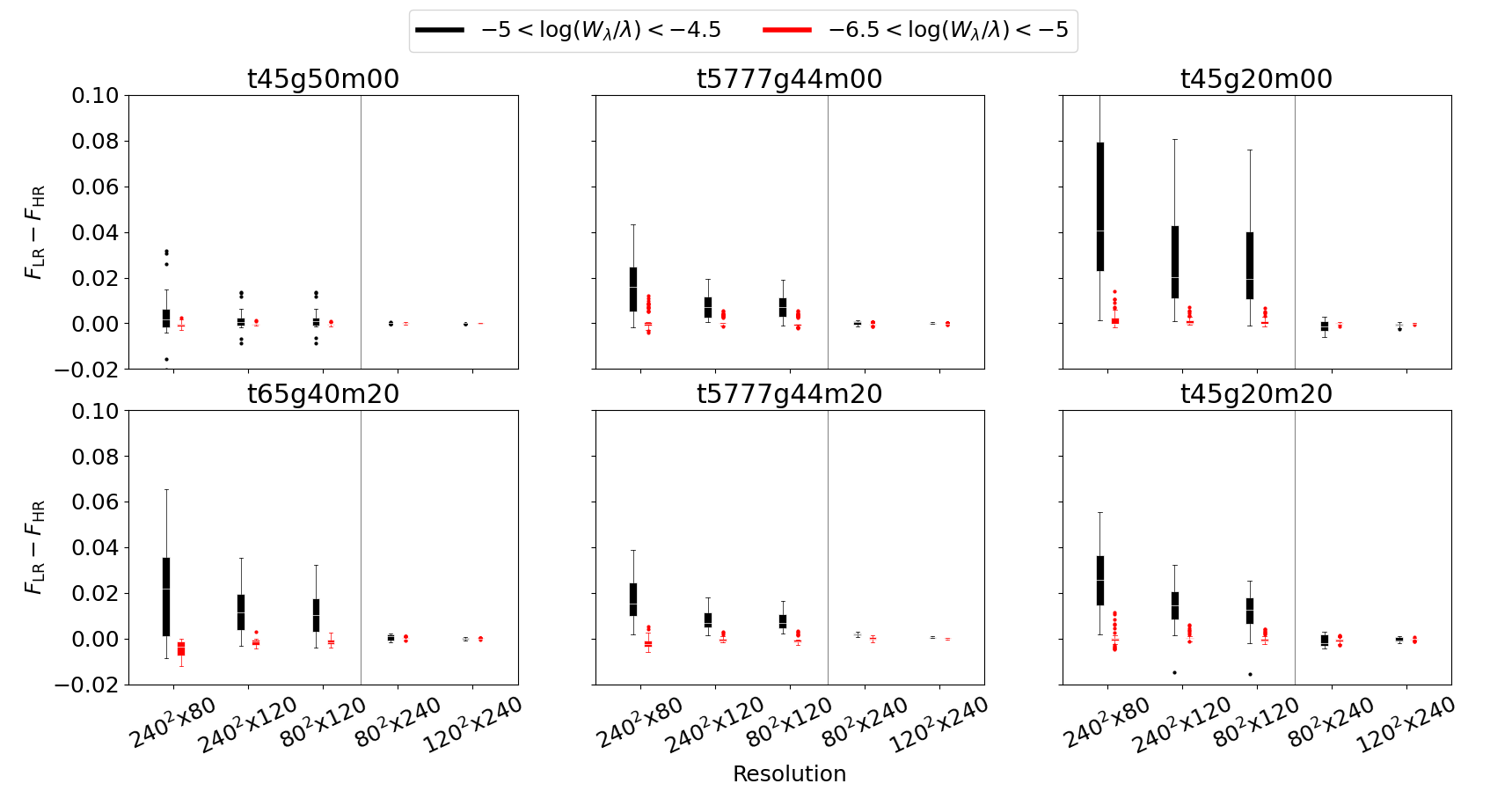}
    
    \caption{Overview of the maximum difference in normalized flux, over the complete line profile, for five different spatial resolutions compared to the reference $240^2\times240$ resolution. Each boxplot contains information from all lines that satisfy the reduced equivalent width requirement shown in the legend. The median of the boxplot is represented by a horizontal white line and outliers with a black or red dot. The vertical grey line divides the models with reduced vertical and horizontal resolution.}
    \label{fig:LineShapeMax_boxplots}
\end{figure*}

\begin{figure*}[h]
    \centering
    \includegraphics[width=0.97\textwidth]{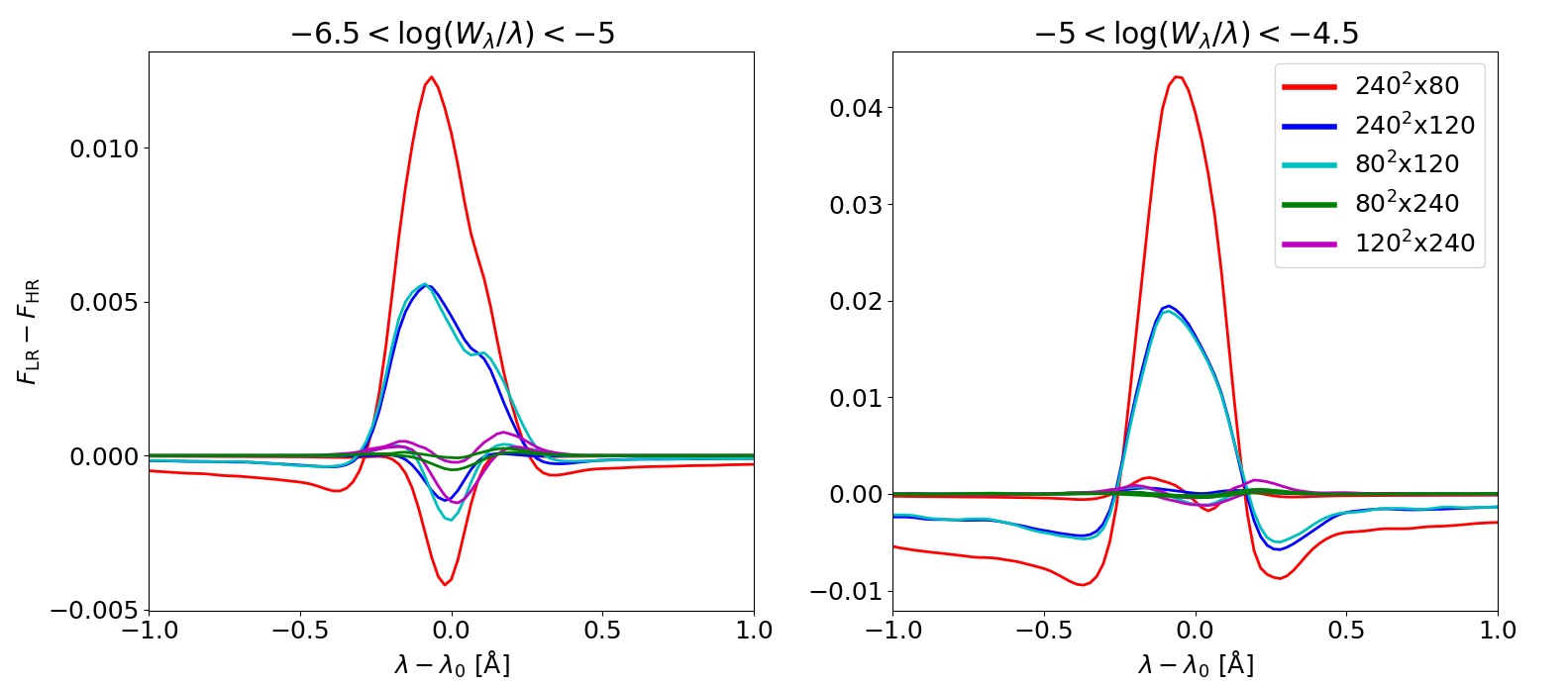}
    
    \caption{Overview of the relative difference in normalized line profile between the lower resolution solar models and the reference high-resolution solar model for two weak and two saturated fictitious Fe I lines with the largest and smallest maximum flux difference, respectively.}
    \label{fig:LineShape_selection}
\end{figure*}

Next, we investigated the impact of downsampling the mesh on the shape of the normalized line profile. Quantifying the change in line shape is of interest when fitting weak or heavily blended spectral lines, or in the case of transmission spectroscopy when trying to constrain the presence of planetary spectral features \citep{Canocchi2023}. In addition, constraining the Li isotopic ratio relies critically on the shape of the Li 670.8 nm line itself \citep{Cayrel2007,Wang2022}. First, the maximum difference in normalized flux, over the complete line profile, between the downsampled and reference model atmosphere is shown in \fig{fig:LineShapeMax_boxplots} for the same lines and model atmospheres discussed above. In this case, the flux difference increases with increasing stellar effective temperature, metallicity and decreasing surface gravity. Similar to the equivalent widths, stronger lines show larger flux errors compared to weak lines across the stellar parameter space. Moreover, the line profile of weak lines are minimally affected by both the vertical and horizontal resolution with maximum differences in normalised flux less than $0.02$. In the case of saturated lines, downsampling the vertical mesh to 120 grid points leads to differences in normalised flux less than $0.08$ for t45g20m00 and less than $0.03$ for all other models. Downsampling the horizontal mesh has minimal impact, with typical differences in normalised flux less than $0.005$ for all model atmospheres. Secondly, the impact of resolution on the shape of the line profile itself is shown in \fig{fig:LineShape_selection} for two weak and two saturated fictitious Fe I lines with the largest and smallest maximum flux difference, respectively. Downsampling the vertical mesh (red, blue and cyan lines) results in a weaker line core and stronger wings. 

From these results, we conclude that, at least for typical applications in stellar spectroscopy, one can safely downsample the horizontal mesh of 3D model atmospheres to $80^2\times240$, reducing the computational cost by a factor nine, introducing abundance errors of $<0.007~\dex$ and differences in normalised flux of $<0.005$. Downsampling both the horizontal and vertical mesh of the model atmosphere to $80^2\times120$ can be reasonable for certain choices of stellar parameters and spectral lines, such as weak lines in cool dwarfs. However, this only results in an additional factor 2 decrease of the computational cost. 

\subsection{Temporal sampling}\label{sec:temporalsampling}
3D simulations only sample a small part of the stellar surface. For example, the stagger solar simulation has a horizontal extent of $8~\mathrm{Mm}$ corresponding to $0.004\%$ of the total surface area. However, stars are observed as point sources and spectral diagnostics are therefore averages over the full stellar disk. Therefore, to accurately model spectral lines, it is necessary to perform spectral synthesis for multiple `statistically independent' snapshots of the simulation, to sample multiple uncorrelated surface granulation patterns of the simulation \citep{Thygesen2016}. In practise, this comes down to performing spectral line calculations for a limited number of snapshots and averaging the resulting line profiles, assuming that the average taken over the sampled snapshots is equivalent to an average over the full stellar disk \citep{Thygesen2017}. The question then becomes, how many snapshots are necessary and how are these selected?

First tests performed by \cite{Asplund2000b,Asplund2000c} in 3D LTE consisted of computing synthetic line profiles every 30 s for a `short' time series of 10 min solar time. This approach resulted in an abundance error of less than $0.02~\dex$ compared to the full time series. 

Later studies expanded on the temporal sampling, for example by spreading snapshots sufficiently in time such that they can be assumed to be statistically uncorrelated \citep{Kucinskas2013,Thygesen2016} while simultaneously requiring that the statistical hydrodynamic properties (temperature, density, velocity) of the subsample resemble that of the full time series \citep{Ludwig2009, Kucinskas2013,Steffen2015}. Subsequent tests by \cite{Nordlander2017} and \cite{Wang2021} in 3D LTE and 3D non-LTE, using equivalent widths, report an abundance error between snapshots of less than $0.03~\dex$ and $0.02~\dex$, respectively. Other temporal sampling tests include the estimation of effective temperature using Balmer lines \citep{Amarsi2018}, finding that five snapshots are sufficient to obtain a precision of 10 K. \cite{Wang2022} studied Li, Fe and a K lines in three metal-poor dwarf stars, finding that the variation of radial velocity between snapshots in 3D non-LTE is at most $150~\mathrm{m/s}$ and in the best case $7~\mathrm{m/s}$. Besides these tests, we can summarize the temporal sampling methods used in previous literature work on 3D (non-)LTE spectrum synthesis. Studies with CO5BOLD models, from the CIFIST grid, consistently aim to use 20 snapshots in the analysis \citep{Ludwig2009,Ludwig2009b,GonzalezHernandez2010,Bonifacio2010,Caffau2011b,Dobrovolskas2012,Dobrovolskas2013,Kucinskas2013,SiqueiraMello2013,Klevas2013,Prakapavicius2013,Dobrovolskas2014,Dobrovolskas2015,Gallagher2015,Steffen2015,Gallagher2016,Thygesen2016,Thygesen2017,Prakapavicius2017,Cerniauskas2017,Mott2017,Harutyunyan2018}. On the other hand, studies with \stagger{} models, usually use between three to five snapshots when performing 3D non-LTE spectrum synthesis \citep{Lind2013,Amarsi2015,Amarsi2016,Amarsi2017,Nordlander2017,Lind2017,Amarsi2018,Bergemann2019,Amarsi2019,Amarsi2020,Wang2021,Wang2022,Lagae2023} or up to 50 snapshots for 3D LTE spectrum synthesis \citep{Collet2018,Amarsi2021}.
For the purpose of selecting snapshots in the \stagger{} grid, we developed a selection method in Python that randomly selects a subsample of $N_t$ snapshots of which the mean hydrodynamic structure is consistent with that of the complete time series, based on earlier work by \cite{Ludwig2009}, \cite{Magic2013b}, \cite{Kucinskas2013} and \cite{Steffen2015}. Specifically, we select a subsample for which the mean vertical stratification, on iso-tau surfaces, of temperature, density and root-mean-square (rms) velocity, and mean of the bolometric surface intensity distribution is consistent with that of the full time series. Here the rms velocity is calculated as follows:
\begin{equation}
    u_\mathrm{rms} = \sqrt{( u^2_x + u^2_y +u^2_z )/3}~~.
\end{equation}

The mean depth-dependent quantities are first computed for each individual snapshot on horizontal layers of equal optical depth at 500 nm in the range $-5<\log\tau_{500}<5$:
\begin{equation}
    \langle X \rangle_{\tau_{500},t} = \frac{1}{N_xN_y}\sum^{N_x}_{x=1}\sum^{N_y}_{y=1}X_{xy,\tau_{500},t}~~,
\end{equation}
with
\begin{equation}
    \tau_{500}=\int(\rho\kappa_{500})\mathrm{d}z~~,
\end{equation}
and $N_x,~N_y$ the number of horizontal grid points. From now on we will simply use $\tau$ to refer to $\tau_{500}$.
Subsequently, the mean quantities are averaged over all snapshots in the subsample \cite[Section 2]{Magic2013b}:
\begin{equation}
    \langle X \rangle_{\tau}^\mathrm{sub} = \frac{1}{N_t}\sum^{N_t}_{t=1}\langle X \rangle_{\tau,t}~~.
\end{equation}
A subsample is deemed acceptable if its mean temperature, density, rms velocity all fall within one standard deviation $\sigma_\tau^{\mathrm{tot}}$ from the mean quantities of the complete time series:
\begin{equation}
    \mid\langle X\rangle_\tau^\mathrm{sub} - \langle X\rangle_\tau^\mathrm{tot}\mid < \sigma_\tau^{\mathrm{tot}}\langle X\rangle~~.
\end{equation}
Here the standard deviation is calculated as follows:
\begin{equation}
    \sigma_\tau^{\mathrm{tot}}\langle X\rangle = \sqrt{\sum^{N_\mathrm{tot}}_{t=1}\Big( \langle X\rangle_{\tau,t}  - \langle X\rangle_\tau^\mathrm{tot}\Big)^2 / N_\mathrm{tot}}~~,
\end{equation}
with the sum extending over the full time series. Besides, an additional constraint is placed on each individual snapshot in the subsample to exclude statistical outliers:
\begin{equation}
    \mid\langle X\rangle_\tau^\mathrm{i} - \langle X\rangle_\tau^\mathrm{tot}\mid <2\cdot\sigma_\tau^{\mathrm{tot}}\langle X\rangle~~.
\end{equation}
In addition, the probability density distribution (PDF) of the bolometric surface intensity at disk centre normalized to its mean value is computed for each subsample and compared to the total time series. 

\begin{figure*}
    \centering
    \includegraphics[width=0.47\linewidth]{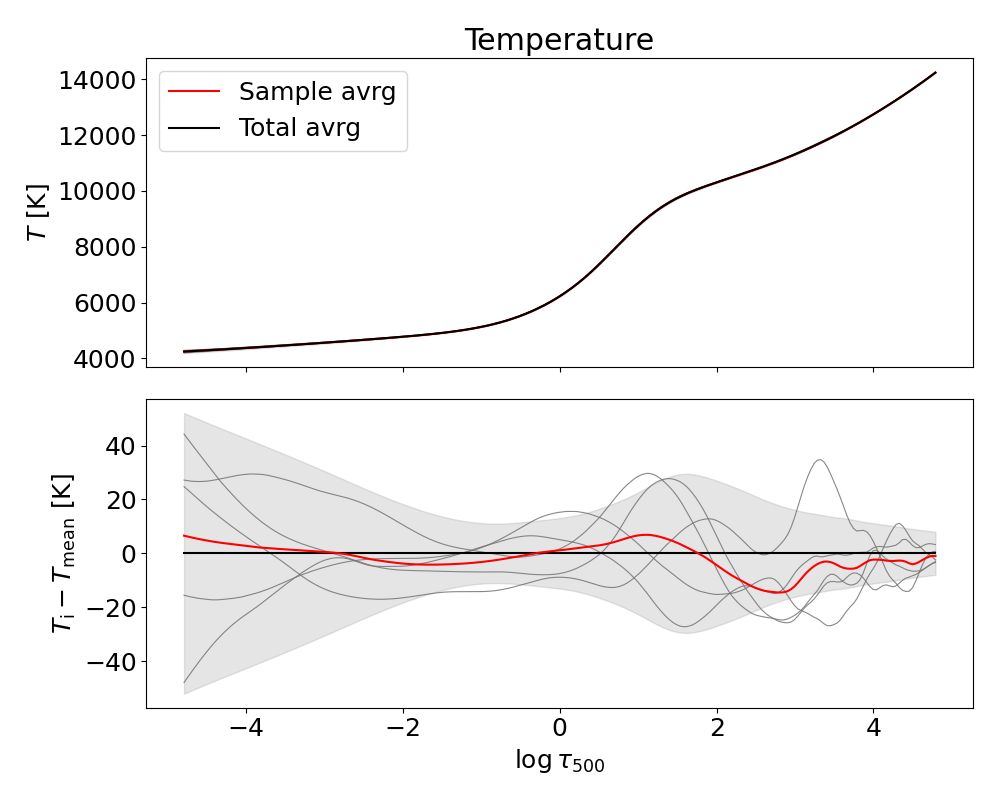}\hfil
    \includegraphics[width=0.47\linewidth]{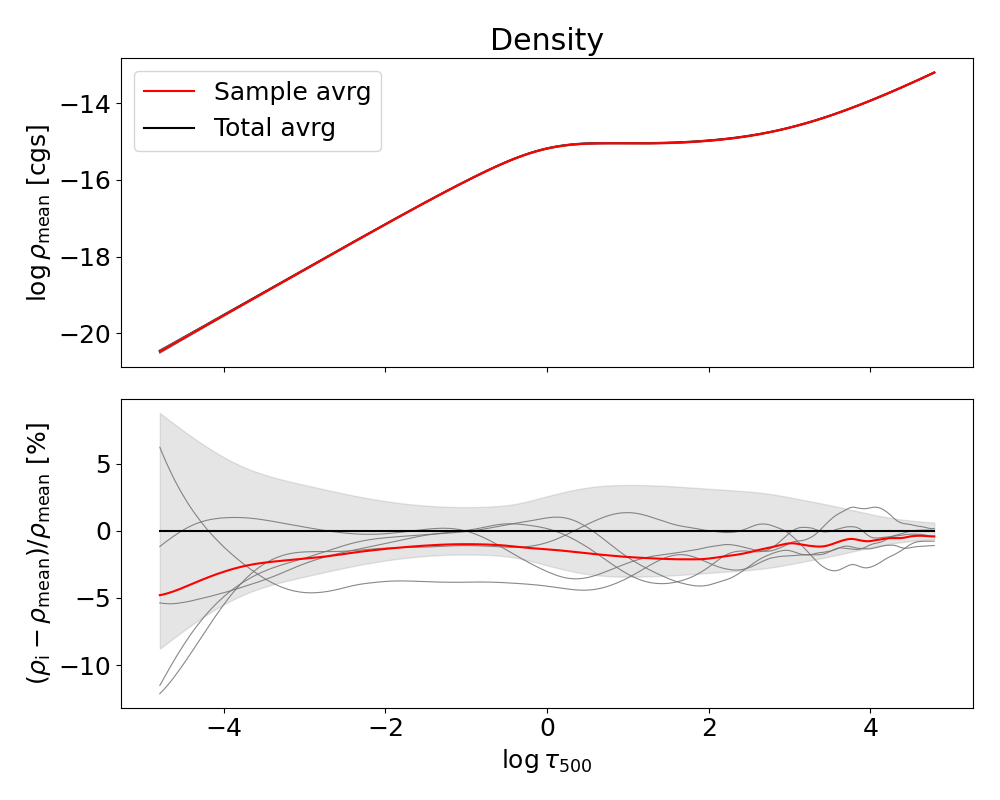}\par\medskip
    \includegraphics[width=0.47\linewidth]{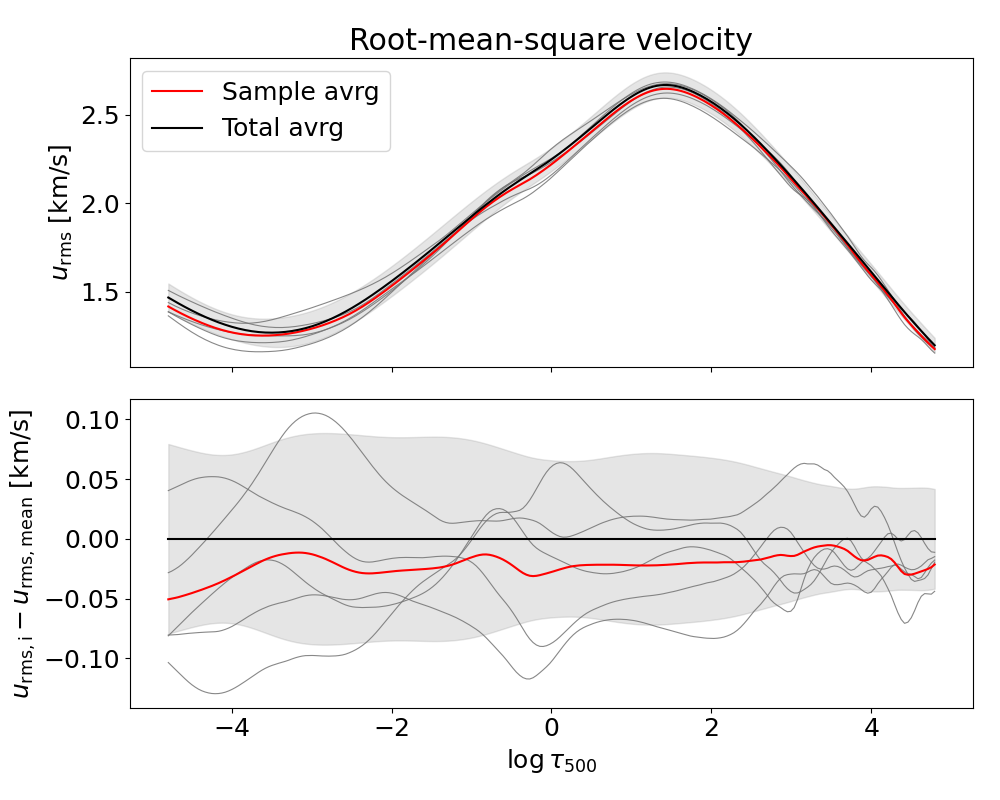}\hfil
    \includegraphics[width=0.47\linewidth]{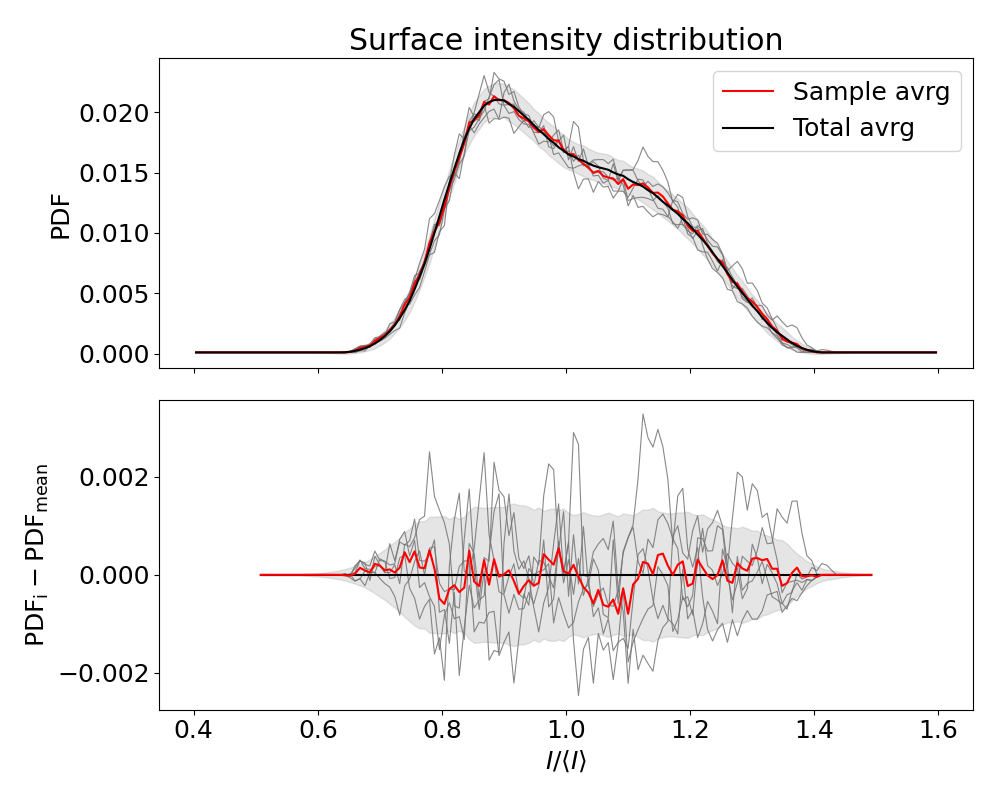}
    \caption{Overview of the subsample selection for the \stagger{} model t5777g44m00 with size $N=5$. Each panel shows the mean of the total time series (black line), its standard deviation (grey shaded area), the mean of the subsample (red line) and individual snapshots (grey lines), for the variables discussed in the text. From top to bottom and left to right, these are temperature, density, rms velocity and the normalized probability distribution of surface intensity.}
    \label{fig:subsample_selection}
\end{figure*}

To make sure that the snapshots in each subsample (with $N_t>3$) are sampled at different times, to assure they are statistically uncorrelated, we enforce a time constraint during the snapshot selection process. Specifically, the method requires that there is at least one snapshot chosen in the following intervals: $t<1/3~ t_\mathrm{tot}$, $1/3~t_\mathrm{tot}<t<2/3\cdot t_\mathrm{tot}$ and $t>2/3~t_\mathrm{tot}$, with $t_\mathrm{tot}$ the total time of the time series. From numerous test runs we conclude that this simple conditions proves sufficient to spread snapshots in time. An example of this selection procedure is shown in \fig{fig:subsample_selection} for the \stagger{} solar model and a subsample consisting of five snapshots $N_t=5$. The mean hydrodynamic quantities from individual snapshots are allowed to deviate from the total mean, but the subsample mean is consistent with the total mean.

Besides the variables mentioned previously, also the amplitude of variations around the mean for temperature, density and rms velocity at every depth layer are calculated:
\begin{equation}
    \delta X_\mathrm{rms}(\tau) = \sqrt{\sum^{N_xN_y}_{i=1}\big(X_i - \langle X \rangle_{\tau}\big)^2/\big(N_xN_y\langle X \rangle_{\tau}^2\big)}
\end{equation}
where the sum is over all grid points $N_x,~N_y$ in a single layer with $\langle X \rangle_{\tau}$ the mean quantity of that layer. These variables are not taken into account in the selection method but serve as an additional tool to review the selected snapshots.

To investigate how many snapshots are necessary to accurately model spectral diagnostics, we selected five subsamples of the solar model (t5777g44m00) consisting of 2, 3, 5, 10 and 20 snapshots respectively, using the selection method discussed previously. For each snapshot in these subsamples, we computed a grid of synthetic spectra, similar to \sect{sec:Hrefinement}. Subsequently, the mean line profile and equivalent width of a subsample were computed 
by averaging the continuum and line flux separately before normalization:
\begin{equation}
    F_\mathrm{subsample}=\frac{\sum_{i=0}^N F_{l,i} }{ \sum_{i=0}^N F_{c,i} }~~,
\end{equation}
with $N$ the number of snapshots in the subsample and $F$ normalized flux. In the following comparison, we treated the mean normalized flux and equivalent width of all subsamples combined, equal to a total of 40 snapshots, as the reference value.

\begin{figure*}
    \centering
    \includegraphics[width=0.46\textwidth]{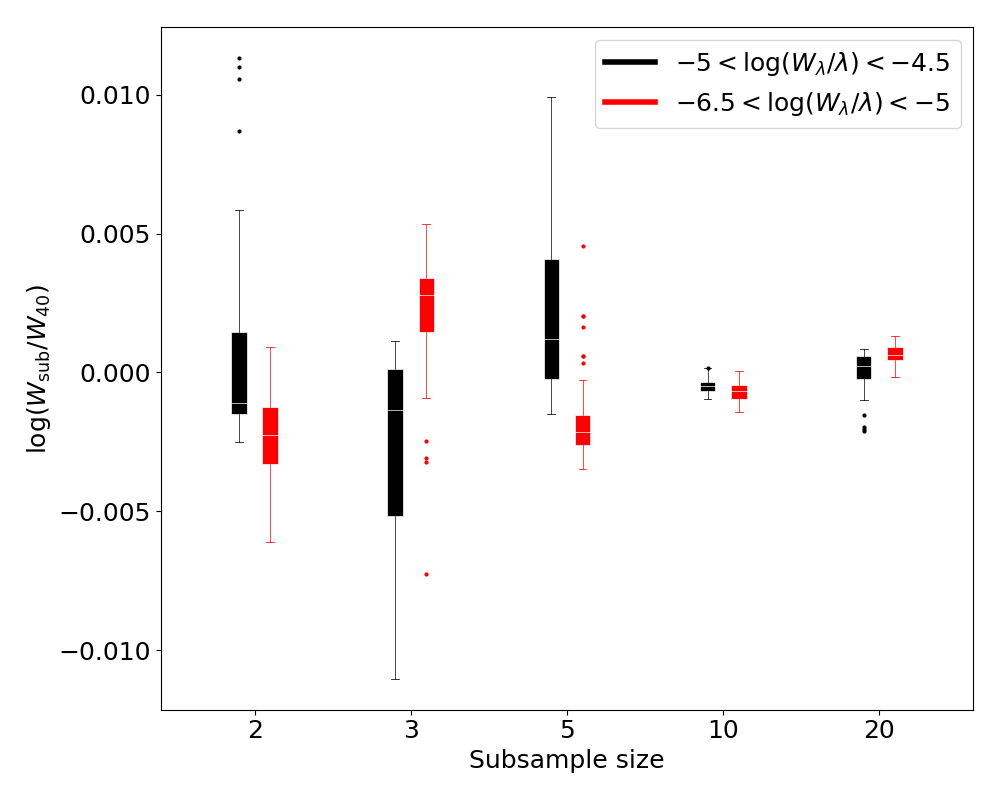}
    \includegraphics[width=0.46\textwidth]{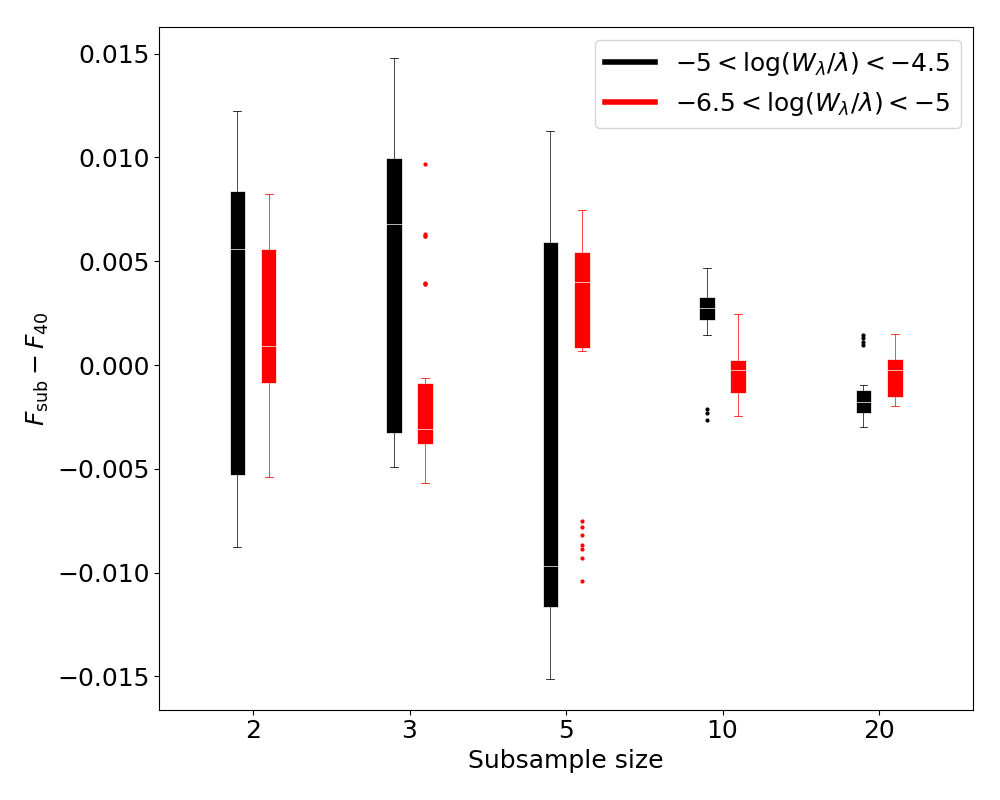}
    \includegraphics[width=0.95\textwidth]{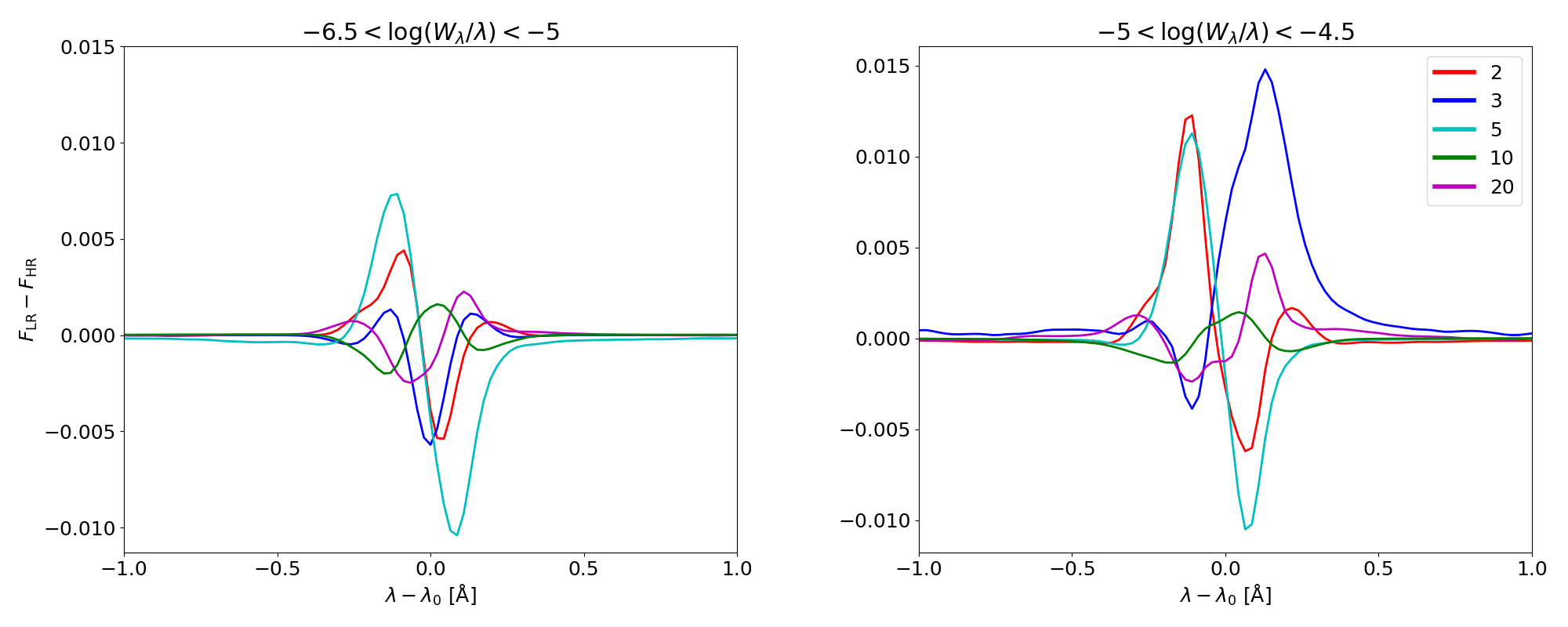}

    \caption{Overview of the relative difference in reduced equivalent width and (maximum) difference in normalized flux for five subsamples compared to the reference sample ($N=40$). Each boxplot contains information from all lines that satisfy the reduced equivalent width requirement shown in the legend. The bottom two panels show the relative difference in normalized line flux profile between the subsamples and the reference sample for two weak and two saturated fictitious Fe I lines with the largest and smallest maximum flux difference, respectively.}
    \label{fig:tempsampling_combined}
\end{figure*}

\fig{fig:tempsampling_combined} compares the equivalent width and normalized line flux of the subsample to that of the reference sample. Once more, we remind the reader that for saturated lines the abundance error has to be multiplied by a factor five, see Eq. \ref{eq:satlines}. Subsamples of size $N<10$ lead to a maximum abundance error of $\approx5\times0.01~\dex$ for saturated lines and $\approx0.005~\dex$ for weak lines. Similarly, these subsample sizes show variations up to $0.015$ in the normalized flux of the line compared to the reference spectral line. We conclude that performing abundance analysis using equivalent widths for weak lines, with at least two suitably-selected snapshots, seems sufficient to reach an accuracy smaller than $0.01~\dex$. On the other hand, to perform accurate spectral fitting it is advised to use at least ten snapshots corresponding to a maximum difference in normalised flux of $0.005$. For example, such accuracy is necessary to accurately fit the ${}^6\mathrm{Li}$/${}^7\mathrm{Li}$ line at $6707~\AA$ which requires residual flux errors lower than $\approx0.005$ \citep{Mott2017, Wang2022}.

\section{Conclusion}
\label{conclusion}
We present an updated version of the \stagger{}-grid that was originally developed by \cite{Magic2013}, by improving the process of scaling, running, and relaxing 3D stellar atmosphere models. The majority of model atmospheres that suffered relaxation problems have been corrected and new dwarf models have been added at intermediate grid points in light of upcoming surveys such as PLATO. This brings the \stagger{}-grid to a total of 243 models, excluding all models with $\feh$ = -4.00 from the original grid, spanning a wide range in $T_\mathrm{eff}$, $\log g$ and metallicity, and covering a large fraction of late-type stars. 

Besides updating the \stagger{}-grid, we have extensively tested the impact of post-processing procedures such as spatial and temporal sampling, on spectroscopic signatures across a six-dimensional parameter space ($T_\mathrm{eff}$, $\log g$ and $\mathrm{[Fe/H]}$, $\log gf$, $E_\mathrm{l}$, $\lambda$). These procedures are necessary to make spectrum synthesis feasible in 3D non-LTE. Hence, it is important to characterize their corresponding systematic errors, providing a theoretical basis for future studies. We find that in general, stronger lines in warm evolved stars are the most sensitive to the spatial resolution of model atmospheres. Downsampling the horizontal mesh from $240^2$ to $80^2$ grid points introduces minimal errors on both abundances $<0.007~\dex$ and fluxes $<0.005$, while reducing the computational cost by a factor nine. Additionally, we find that already two snapshots are sufficient to perform an abundance analysis with equivalent width, introducing a maximum error of $0.01~\dex$. On the other hand, we advise using at least 10 snapshots when trying to accurately fit line profiles, corresponding to a maximum flux error of $0.005$. We remind the reader that these tests have been performed in 3D LTE, although we do expect these results to be approximately valid also in 3D non-LTE. Nevertheless, future studies performing 3D non-LTE spectrum synthesis should investigate the impact of downsampling and temporal sampling for their specific case. 

In light of these results, we publicly release ten snapshots for all model atmospheres except for those with $\feh$ = -4.00, both in their original format and a downsampled format with mesh dimensions $80^2\times240$, occupying 316 MB and 35 MB of memory per snapshot for the two formats, respectively. These snapshots will contribute significantly to several applications in the astronomy community, from spectroscopic studies, abundance determinations, to stellar characterization. With upcoming space missions, such as PLATO, the more refined grid will provide even more possibilities for different studies. 

\begin{acknowledgements}
We thank the anonymous referee for their constructive feedback, which has improved the manuscript. In addition, we thank H. G. Ludwig and B. Freytag for many fruitful discussions and insights on acoustic wave heating in 3D model atmospheres. The numerical results presented in this work were obtained at the Centre for Scientific Computing, Aarhus: http://phys.au.dk/forskning/cscaa/. In addition, computational resources were used from the Swedish National Infrastructure for Computing (SNIC) at UPPMAX and the PDC Center for High Performance Computing, partially funded by the Swedish Research Council through grant agreement no. 2018-05973. AMA acknowledges support from the Swedish Research Council (VR 2020-03940). CL and KL acknowledges funds from the European Research Council (ERC) under the European Union’s Horizon 2020 research and innovation program (Grant Agreement No. 852977). RT acknowledges funding by NASA grants 80NSSC20K0543 and 80NSSC22K0829. This work was supported by the Ministry of Education, Youth and Sports of the Czech Republic through the e-INFRA CZ (ID:90254). This research was supported by computational resources provided by the Australian Government through the National Computational Infrastructure (NCI) under the National Computational Merit Allocation Scheme and the ANU Merit Allocation Scheme (project y89).

\end{acknowledgements}
\bibliographystyle{aa_url} 
\bibliography{main.bib}


\end{document}